\documentclass[twocolumn]{aa}
\input epsf
\usepackage{graphicx}
\begin{document}

%
   \title{New abundances of planetary nebulae in the Galactic
Bulge\thanks{Based
on observations made at the European Southern Observatory (Chile) and Laborat\'orio
Nacional de Astrof\'\i sica (Brasil)}}

   \author{A.V. Escudero,
          \inst{1}
          R.D.D. Costa
          \inst{1}
          \and
          W.J. Maciel
          \inst{1}
          }

   \offprints{A.V. Escudero}

   \institute{Departamento de Astronomia, IAG/USP, C.P. 3386, 01060-970
S\~ao Paulo - SP, Brasil\\
\email{escudero@astro.iag.usp.br; 
roberto@astro.iag.usp.br; \\
maciel@astro.iag.usp.br}
}

   \date{Received 2 June 2003 / Accepted 29 September 2003 }

   \authorrunning{A.V. Escudero et al.}
   \titlerunning{Abundances of PNe towards the Bulge}
  
   \abstract{
New observations and derived chemical abundances are reported for a sample of 57
bulge planetary nebulae (PN). Together with our previous results, a total of over
a hundred objects have been analyzed, which constitute one of the largest samples
of bulge nebulae studied under homogeneous conditions, including equipment and
reduction procedures. In general, our data show a good agreement with some recent
results in the literature, in the sense that the average abundances of bulge PN are
similar to those from disk objects, however showing a larger dispersion.

      \keywords{planetary nebulae --
                galactic bulge --
                spectroscopy
                }
  }
 \maketitle

%

\section{Introduction}

In the past few years, many papers have been published
dealing with the kinematics and abundances of the galactic bulge.
Most of these works on bulge abundances are concerned with heavy
elements produced by supernovae, so that light elements such as
helium and nitrogen have had a smaller share of attention.

Planetary nebulae (PN) constitute an important tool in the study of
the chemical evolution of the bulge, providing accurate
determinations of the abundances of light elements produced by
progenitor stars of different masses. In fact, PN offer the
possibility of studying both light elements produced in low
mass stars, such as He and N, and also heavier elements which
result from the nucleosynthesis of large mass stars, such as
oxygen, sulfur and neon, which are present in the interstellar
medium at the stellar progenitor formation epoch.

Previous determinations of chemical abundances of bulge PN have
shown that, on average, these objects have abundances similar
to disk planetary nebulae (Escudero \& Costa 2001, Liu et al. 2001,
Cuisinier et al. 2000, Costa \& Maciel 1999, Ratag et al. 1997). However,
the total number of bulge PN with accurate abundances is still
small; furthermore, a correlation between the chemical abundances
and bulge kinematics is still to be determined, in contrast with
the observed properties of the galactic disk.

In the present paper, we report new observations and derive chemical
abundances for a sample of  57 bulge PN. These results are compared
with previous data from our own group and other groups as well, both
regarding the galactic bulge and other galactic systems, such as the
galactic disk and halo. 

In section 2 we present our new observations for a sample of bulge PN
and comment on the reduction procedure. In section 3 we present our method
to derive the chemical abundances. A discussion of the main uncertainties
of the physical parameters is given in section 4. A comparison of our
results with published data is given in section 5, and in section 6 we
discuss our results an present our main conclusions.


\section{Observations and data reduction}

\subsection{Observations}

In this work we present results for 57 new PNe,
extending and completing a previous paper (Escudero \& Costa 2001),
in which we have presented our results for a sample of
45 objects toward the galactic bulge.

The objects were selected
according to their distances, taken from the literature (Schneider
\& Buckley 1996, van de Steene \& Zijlstra 1995 and Zhang 1995), adopting
as probable bulge objects those with heliocentric distances greater than
5 kpc. In spite of the sometimes larger uncertainties  for these distances,
this is still the best homogeneous criterion to define the sample.
Other criteria like diameters, H$\beta$ or radio fluxes are even less reliable.
It should be noted that in this paper we did not use the same criterion
used by Escudero \& Costa (2001), where all objects towards the bulge from
Beaulieu et al. (1999) and Kohoutek (1994) were observed. These
objects do not have distance determinations, but as most of them have
diameters smaller than 20", they should be at or near the bulge according to
the criteria of Gathier et al. (1983). 

Adopted distances are listed in table
3. We divided our sample in two samples: those with heliocentric distances
greater than 5 kpc (Sample 1), and those with heliocentric distancies smaller
than 5 kpc or unknown (Sample 2). Figure 1 displays the distance distribution
for our Sample 1 objects, combined with objects chosen from the literature
(see section 5 for detailed references). Adopting these distances and the
criteria mentioned above we would expect some 80 \% of the sample to be in
the galactic bulge, a fraction similar to the estimates given by Pottasch (1990),
to which the reader is refered for a detailed discussion on the characterization
of galactic bulge PN.

\bigskip
\begin{figure}
\epsfxsize=230pt \epsfbox{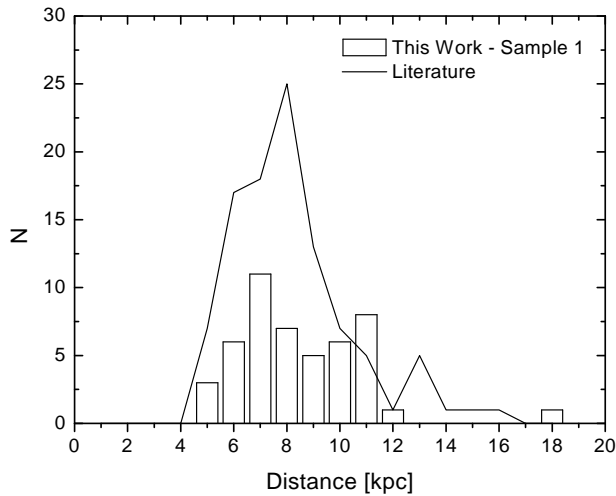}
\caption[]{Distance distribution for our sample a (objects with
heliocentric distances greater than 5 kpc) and objects from
the literature.
}
\end{figure}
%

The program objects were observed with two telescopes: 1.60m LNA
(Laborat\'orio Nacional de Astrof\'\i sica, Bras\'opolis - Brasil),
and 1.52m ESO (European Southern Observatory). Each object was
observed at least twice in order to secure good final
 spectra. We selected the exposure time for
each object in order to have a good S/N in the region of the
H$\gamma$ line, due to the importance of the [OIII]$\lambda$4363
line in the temperature diagnostics.

For all objects, we used the long slit of 2 arcsecs width.
At LNA we used a 300 l/mm grid with 4.4 A/pix dispersion, and at ESO
the used grid was 600 l/mm with 2.4 A/pix dispersion.
Each night at least three spectrophotometric standard
stars were observed to improve the flux calibration.

Table 1 list the PN G designation (first column), the 
usual names of the observed objects (column 2), equatorial coordinates for epoch 2000
(columns 3 and 4), date of observation (column 5), observatory (column 6)
and the exposure time in seconds (last column).
For some objects, spectra were taken with different exposure times:
a short exposure to measure the Balmer series lines (and therefore to derive interstellar
extinction), and a longer exposure (usually with H$\alpha$ saturated)
to measure the fluxes of lower intensity lines.

Table 2 (available electronically) display the line fluxes of the
reddened lines. For the cases where the line blend [OII]$\lambda\lambda$7319,7330
was impossible do separate, their intensities are added in the
[OII]$\lambda$7319 column.

Figure 2 shows typical spectra from our sample. The upper panel shows M 3-26, an
object with well defined lines, and the other shows KFL 4,
a poorer object, observed with lower resolution.

\begin{table*}
\caption[]{Log of the observations}
\begin{flushleft}
\begin{tabular}{lllllll}
\hline\noalign{\smallskip}
PN G & Name & RA(2000) & DEC (2000) & Date of Obs. & Place & Exp. Time (s)\\
\noalign{\smallskip}
\hline\noalign{\smallskip}
000.4-02.9 &M 3-19 & 17 58 18 & - 30 00 10 & 19-may-01 & ESO & 2x 900\\
000.4-01.9 &M 2-20 & 17 54 25 & - 29 36 09 & 19-may-01 & ESO & 1x 600\\
000.7+04.7 &H 2-11 & 17 29 26 & - 25 49 06 & 11-jun-02 & LNA & 2x 1800\\
000.9-02.0 &Bl 3-13 & 17 56 02 & - 29 11 15 & 22-may/11-jul-01 & ESO & 2x 900, 2x 2400\\
001.2+02.1 &He 2-262 & 17 40 15 & - 26 43 21 & 19-may-01 & ESO & 2x 1200\\
001.7+05.7 &H 1-14 & 17 28 02 & - 24 25 28 & 12-jul-01 & ESO & 2x 1800\\
002.0-02.0 &H 1-45 & 17 58 23 & - 28 14 54 & 19-may-01 & ESO & 3x 600\\
002.2-06.3 &H 1-63 & 18 16 19 & - 30 07 36 & 13-jul-01 & ESO & 2x 1200\\
002.2-02.5 &KFL 2 & 18 01 00 & - 28 16 19 & 22-may-01 & ESO & 2x 1800\\
002.6+08.2 &H 1-11 & 17 21 18 & - 22 18 36 & 14-jul-01 & ESO & 2x 1200\\
003.0-02.6 &KFL 4 & 18 02 52 & - 27 41 02 & 22-may/15-jul-01 & ESO & 2x 1200, 1x 3600, 1x 1800\\
003.3-07.5 &KFL 19 & 18 23 09 & - 29 43 25 & 12-jul-01 & ESO & 2x 1800\\
003.6-02.3 &M 2-26 & 18 03 12 & - 26 58 31 & 22-may-01 & ESO & 2x 1200\\
004.0-05.8 &Pe 1-12 & 18 17 43 & - 28 17 24 & 13-jul-01 & ESO & 1x 1200, 1x 1800\\
004.8-05.0 &M 3-26 & 18 16 09 & - 27 14 50 & 12-jul-01 & ESO & 2x 900\\
004.8+02.0 &H 2-25 & 17 49 02 & - 23 42 57 & 19/21-may-01 & ESO & 1x 900, 1x 1200\\
005.5+02.7 &H 1-34 & 17 48 08 & - 22 46 48 & 21-may-01 & ESO & 2x 1200\\
005.7-05.3 &M 2-38 & 18 19 26 & - 26 35 20 & 13-jul-01 & ESO & 2x 120\\
006.4+02.0 &M 1-31 & 17 52 41 & - 22 21 57 & 19-may-01 & ESO & 1x 600, 2x 300\\
006.8+02.3 &Th 4-7 & 17 52 21 & - 21 51 12 & 19-may-01 & ESO & 2x 900\\
007.0-06.8 &Vy 2-1 & 18 27 59 & - 26 06 48 & 13-jun-01 & LNA & 2x 300\\
008.1-04.7 &M 2-39 & 18 22 02 & - 24 10 41 & 11-jun-02 & LNA & 2x 300\\
008.2+06.8 &He 2-260 & 17 38 57 & - 18 17 35 & 12-jul-01 & ESO & 1x 1200, 1x 600\\
008.8+05.2 &Th 4-2 & 17 46 09 & - 18 39 33 & 12-jul-01 & ESO & 2x 1800\\
009.4-09.8 &M 3-32 & 18 44 43 & - 25 21 34 & 12-jul-01 & ESO & 2x 900\\
009.6-10.6 &M 3-33 & 18 48 13 & - 25 28 56 & 13-jul-01 & ESO & 2x 600\\
009.8-04.6 &H 1-67 & 18 25 06 & - 22 34 51 & 11-jun-02 & LNA & 2x 900\\
010.7-06.7 &Pe 1-13 & 18 34 53 & - 22 43 18 & 13-jun-01 & LNA & 1x 600, 2x 1200\\
011.0+06.2 &M 2-15 & 17 46 54 & - 16 17 27 & 12-jul-01 & ESO & 2x 900\\
011.1+11.5 &M 2-13 & 17 28 34 & - 13 26 17 & 13-jul-01  & ESO & 2x 1200\\
011.3-09.4 &H 2-48 & 18 46 35 & - 23 26 41 & 10-jun-02 & LNA & 2x 30, 2x 300\\
013.4-03.9 &M 1-48 & 18 29 30 & - 19 06 51 & 10-jun-02 & LNA & 2x 1200\\
013.7-10.6 &Y-C 2-32 & 18 55 31 & - 21 49 39 & 13-jul-01 & ESO & 2x 600\\
018.6-02.2 &M 3-54 & 18 33 04 & - 13 44 21 & 11-jun-02 & LNA & 2x 1200\\
018.9+03.6 &M 4-8 & 18 12 06 & - 10 43 05 & 11-jun-02 & LNA & 2x 1800\\
018.9+04.1 &M 3-52 & 18 10 30 & - 10 28 60 & 11-jun-02 & LNA & 2x 1800\\
352.0-04.6 &H 1-30 & 17 45 08 & - 38 08 55 & 10-jun-02 & LNA & 2x 1200\\
352.1+05.1 &M 2-8 & 17 05 31 & - 32 32 08 & 13-jun-01 & LNA & 2x 800\\
352.8-00.2 &H 1-13 & 17 28 28 & - 35 07 32 & 22-may-01 & ESO & 2x 1200\\
353.7+06.3 &M 2-7 & 17 05 13 & - 30 32 14 & 11-jul-01 & ESO & 2x 900\\
353.8-01.2 &K 6-3 & 17 35 27 & - 34 47 42 & 14-jul-01 & ESO & 1x 2400, 1x 3600\\
354.6-01.7 &K 6-5 & 17 38 54 & - 34 27 39 & 15-jul-01 & ESO & 2x 3600\\
356.1+02.7 &Th 3-13 & 17 25 19 & - 30 40 44 & 20-may-01 & ESO & 2x 1200\\
356.3-00.3 &Th 3-34 & 17 37 45 & - 32 15 29 & 14-jul-01 & ESO & 1x 3600, 1x 2400\\
356.7-06.4 &H 1-51 & 18 04 30 & - 34 58 00 & 15-jul-01 & ESO & 2x 1800\\
356.7-04.8 &H 1-41 & 17 57 19 & - 34 09 50 & 10-jun-02 & LNA & 2x 1200\\
357.0+02.4 &M 4-4 & 17 28 50 & - 30 07 46 & 21-may-01 & LNA & 2x 1200\\
357.1-04.7 &H 1-43 & 17 58 15 & - 33 47 39 & 11-jun-02 & LNA & 2x 300, 2x 1200\\
357.2-04.5 &H 1-42 & 17 57 25 & - 33 35 44 & 11-jun-02 & LNA & 2x 150\\
358.3-02.5 &Al 2-O & 17 51 45 & - 32 03 04 & 22-may-01 & ESO & 2x 1800\\
358.5-04.2 &H 1-46 & 17 59 02 & - 32 21 44 & 11-jun-02 & LNA & 2x 150, 1x 600\\
358.5-02.5 &M 4-7 & 17 51 45 & - 31 36 00 & 22-may-01 & ESO & 2x 1800\\
358.6+01.8 &M 4-6 & 17 35 14 & - 29 03 11 & 13-jul-01 & ESO & 2x 1800\\
358.8+04.0 &Th 3-15 & 17 27 09 & - 27 44 20 & 11-jun-02 & LNA & 2x 1800\\
359.8+05.6 &M 2-12 & 17 24 01 & - 25 59 23 & 11-jul-01 & ESO & 2x 900\\
359.8+06.9 &M 3-37 & 17 19 13 & - 25 17 15 & 12-jul-01 & ESO & 2x 1800\\
359.9+05.1 &M 3-9 & 17 25 43 & - 26 11 54 & 13-jul-01 & ESO & 2x 1800\\
\noalign{\smallskip}
\hline
\end{tabular}
\end{flushleft}
\end{table*}

\begin{table*}
\caption[]{Line fluxes (available electronically)}
\begin{flushleft}
\begin{tabular}{lrrrrrrrrrr}
\hline\noalign{\smallskip}
Line & Al 2-0 & Bl 3-13 & H1-11 & H1-13 & H1-14 & H1-30 & H1-34 & H1-41 & H1-42 & ...\\
\noalign{\smallskip}
\hline\noalign{\smallskip}
$\rm{[OII]}\lambda$3726+29 & 179.3 & - & 7.4 & - & 48.0 & 83.1 & 57.7 & 31.7 & 18.5 & ...\\
$\rm{[NeIII]}\lambda$3869 & 124.5 & 49.3 & 76.6 & 74.8 & 100.8 & 112.9 & - & 59.4 & 88.5 & ...\\
:\\
\noalign{\smallskip}
\hline
\end{tabular}
\end{flushleft}
\end{table*}

\bigskip
\begin{figure*}
\centering
\includegraphics[width=17cm]{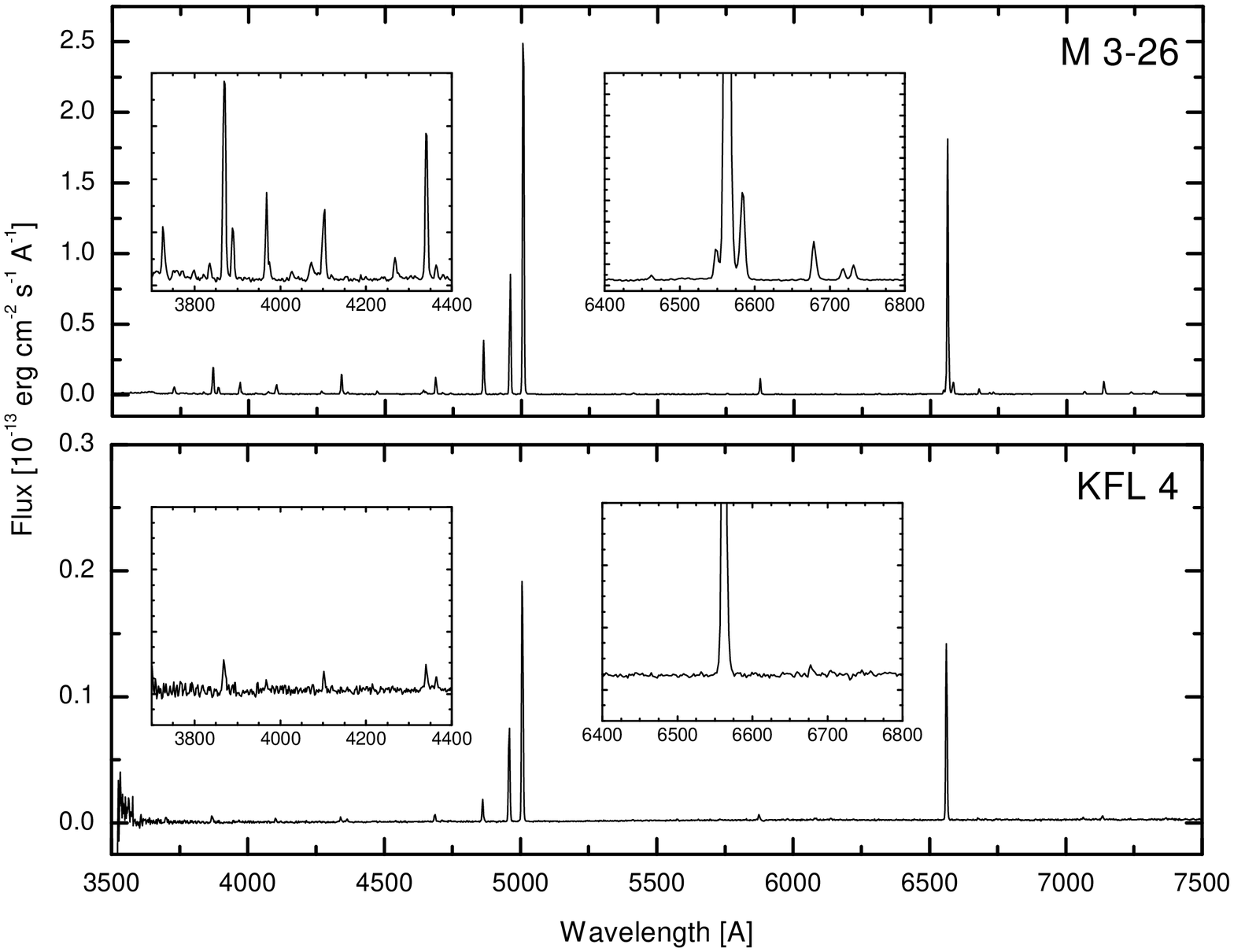}
\caption[]{Typical spectra from our sample. The upper panel shows M 3-26,
with well defined lines, and the lower panel shows KFL 4,
a poorer object.
}
\end{figure*}

\subsection {Interstellar Extinction}

The application of the extinction curve of Fitzpatrick (1999) had better
results than the use of the curve of
Cardelli et al. (1989) (CCM), when we compared the predicted H$\gamma$/H$\beta$=0.47,
(adopting Osterbrock's case B)
with the calculated values after interstellar extinction correction.
A small difference between the two curves in the H$\gamma$ and H$\alpha$ regions
 (see fig. 6 of Fitzpatrick 1999) results
in a significant discrepancy with respect to the recombination values when we use
the CCM curve. Using the Fitzpatrick (1999) curve this difference
is considerably reduced. These results are displayed in figure 3, where both
distributions are plotted. Is easy to see that the distribution of
H$\gamma$/H$\beta$ ratios calculated with the Fitzpatrick curve is narrower
and that its center is closer to the theoretical value of 0.47. 

In order to bring the value of H$\gamma$ closer to the recombination value using
the CCM curve, it would be necessary to use a value of the ratio of total
to selective absorption R$_{V}$ higher than
6.0, which is extremely large and inconsistent with the
value determined by Stasi\'nska et al. (1992).
The main consequence for the poor interstellar correction in the H$\gamma$ region
is the poor determination of the electron temperature for O$^{++}$.
High O[III]$\lambda$4363 values would result in overestimated temperatures
and low abundance values, as shown by other authors (K\"oppen et al. 1991,
Stasi\'nska et al. 1998) who expected that some of the objects lacking
O[III]$\lambda$4363 might be the most metal rich.

In this work we have used the Fitzpatrick (1999) extinction
curve, deriving E(B-V) from the observed Balmer ratio H$\alpha$/H$\beta$
and adopting the theoretical value H$\alpha$/H$\beta$ = 2.85, with R$_{V}$ = 3.1.
\bigskip
\begin{figure}[h]
\epsfxsize=230pt \epsfbox{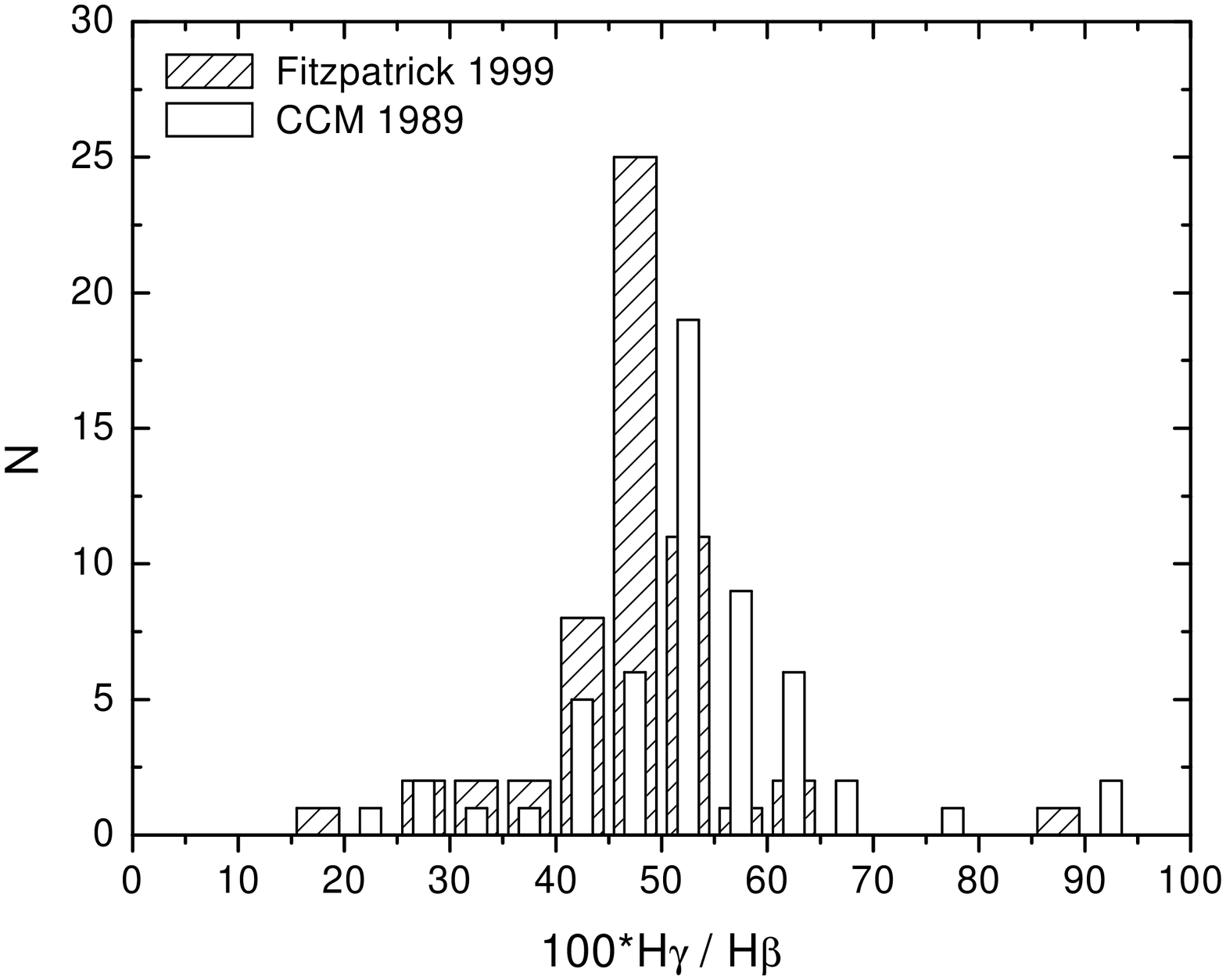}
\caption[]{The distribution of the $H\gamma/H\beta$ values
corrected by the Fitzpatrick (1999) (hatched) and
Cardelli et al. (1989) (white) interstellar extinction curves.
}
\end{figure}
%

Figure 4 displays the distribution of E(B-V) derived
for Sample 1 (see section 2.1) compared with data from the literature and
from Escudero \& Costa (2001). It can be seen that the distribution for the present 
data agrees well with our previous work, and is in reasonable agreement with
data from the literature, which are on average nearer the galactic center.

\bigskip
\begin{figure}[h]
\epsfxsize=230pt \epsfbox{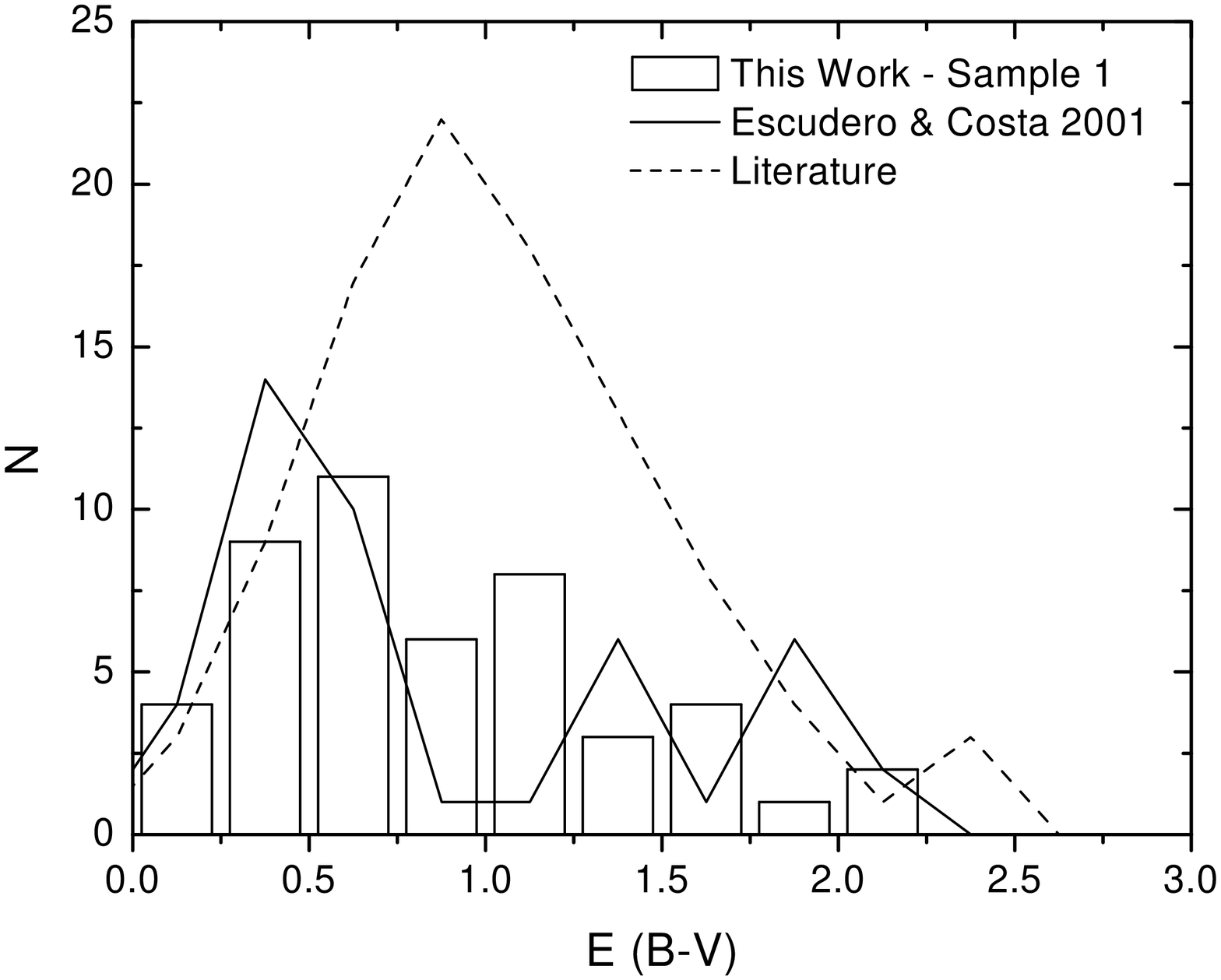}
\caption[]{Distribution of E(B-V) values for our sample.
Also included are the distribution of Escudero \& Costa (2001) 
and data from the literature.
}
\end{figure}
%

\section {Determination of chemical abundances}
\subsection {Physical parameters}

The quality of the physical parameters is crucial for
a good derivation of chemical abundances, as
line emissivities of the ions are extremely sensitive to
the electron temperature and density of the nebulae.

For the determination of electron temperatures we used
the following line ratios for oxygen and nitrogen:
[OIII]$\lambda\lambda$4363,5007 and [NII]$\lambda\lambda$5755,6584,
from which we can obtain estimates of the electron temperature for
high and low ionization regions.
The electron density was derived from
the sulfur line ratio [SII]$\lambda\lambda$6716,6731, giving
therefore an average value for the whole nebula.
Excitation classes were also defined for the object following the Webster (1988)
criterium. For each nebula the H$\beta$ surface flux was calculated
using the derived H$\beta$ flux and the fraction of the nebula within
the slit.  

In table 3 we give the observed (column 2) H$\beta$ intensities
and corrected by interstellar extinction (in erg cm$^{-2}$ s$^{-1}$)
(column 3), diameters in arcsecs from Acker et al. (1991) (column 4),
H$\beta$ surface flux (in erg cm$^{-2}$ s$^{-1}$ arcsec$^{-2}$),
  the interstellar extinction E(B-V)
(column 6), the electron density in 10$^{3}$cm$^{-3}$ (column 7), the electron
temperatures in 10$^{4}$ K (columns 8 and 9), excitation class (column 10) and
heliocentric distances in kpc (column 11).

\begin{table*}
\caption[]{Physical parameters and distances}
\begin{flushleft}
\begin{tabular}{lllllllllll}
\hline\noalign{\smallskip}
PNe & log$_{10}$(H$\beta$)&log$_{10}$(H$\beta$)$_{0}$& Diam & log$_{10}$(SF)& E(B-V) & n[SII] & T[NII] & T[OIII] & EC & Dist\\
\noalign{\smallskip}
\hline\noalign{\smallskip}
Al 2-O & -14.27 & -12.47 & -&-&1.22 & 0.40 & 0.86 & 1.53& 9 &5.69\\
Bl 3-13 & -13.41 & -11.69 & 5.2&-12.71&1.17 & 1.88 & - & 0.81& 3 &6.87\\
H 1-11 & -12.88 & -11.55 & 6.4&-12.66&0.90 & 1.92 & - & 0.93& 5 &6.85\\
H 1-14 & -13.42 & -11.70 & 6.6&-12.82&1.17 & 0.97 & 1.51 & 1.56& 6 &5.59\\
H 1-30 & -13.10 & -11.83 & 5.4&-12.86&0.86 & 6.90 & 0.88 & 1.01& 6 &5.21\\
H 1-34 & -13.33 & -11.55 & -&-&1.20 & 3.55 & 0.72 & 0.9& 0.5 &10.45\\
H 1-41 & -12.36 & -11.88 & 9.6&-13.16&0.33 & 3.51 & 1.11 & 0.96& 6 &5.98\\
H 1-43 & -12.43 & -11.46 & 2.0&-12.06&0.66 & 9.52 & 0.55 & -& 0 &11.13\\
H 1-45 & -13.22 & -11.11 & 6.0&-12.19&1.43 & - & - & 4.47& 7 &7.65\\
H 1-46 & -12.12 & -10.87 & -&-&0.85 & 5.31 & 1.47 & 0.96& 2 &7.29\\
H 1-63 & -11.77 & -11.20 & 7.0&-12.35&0.38 & 20.00 & 2.11 & 1.09& 1 &11.00\\
H 1-67 & -12.64 & -11.83 & 5.6&-12.88&0.55 & 1.46 & 0.99 & 1.06& 7 &7.13\\
H 2-11 & -13.91 & -11.02 & 2.7&-11.75&1.96 & 20.00 & 0.85 & 1.02& 1 &7.90\\
H 2-25 & -13.66 & -11.46 & 4.4&-12.41&1.49 & 0.52 & 0.84 & -& 0.5 &11.40\\
H 2-48 & -11.44 & -10.54 & 2.0&-11.14&0.61 & 13.23 & 1.32 & 1.03& 0.5 &5.49\\
He 2-260 & -12.39 & -11.69 & 10.0&-12.99&0.47 & 20.00 & 1.07 & -& 0.5 &11.72\\
He 2-262 & -13.84 & -11.43 & 4.0&-12.33&1.63 & 4.19 & - & -& 4 &6.66\\
KFL 2 & -14.12 & -13.40 & 5.4&-14.43&0.49 & - & - & 1.70& 9 &11.07\\
KFL 4 & -14.00 & -12.75 & 3.0&-13.53&0.85 & - & - & 1.38& 7 &17.66\\
KFL 19 & -13.49 & -13.28 & 7.8&-14.48&0.14 & - & - & 0.89& 3 &11.38\\
M 1-48 & -12.66 & -11.79 & 4.8&-12.77&0.59 & 1.58 & 0.84 & 0.87& 5 &9.46\\
M 2-7 & -12.84 & -11.86 & 7.8&-13.05&0.66 & 0.83 & 0.75 & -& 1 &7.32\\
M 2-8 & -12.37 & -11.29 & 4.2&-12.21&0.73 & 1.67 & 0.95 & 1.05& 7 &7.15\\
M 2-12 & -12.52 & -11.46 & 5.0&-12.46&0.71 & 8.24 & 0.71 & 0.71& 0 &8.15\\
M 2-13 & -12.73 & -11.77 & 7.0&-12.91&0.66 & 4.65 & 1.00 & 0.91& $>$4 &10.26\\
M 2-15 & -12.65 & -11.61 & 5.8&-12.68&0.70 & 2.28 & 1.47 & 0.94& 6 &6.66\\
M 2-20 & -12.93 & -11.16 & 6.6&-12.29&1.19 & 1.09 & 0.79 & 0.88& 5 &6.08\\
M 2-26 & -13.28 & -11.94 & 9.2&-13.20&0.91 & 0.31 & 0.81 & 0.99& 2 &8.20\\
M 2-38 & -13.06 & -12.97 & 9.3&-14.24&0.07 & 0.89 & 1.57 & 1.21& 9 &7.54\\
M 2-39 & -12.19 & -11.21 & 3.2&-12.02&0.66 & 2.92 & 1.11 & 3.35&3-4&10.05\\
M 3-19 & -13.65 & -12.99 & 5.7&-14.05&0.45 & - & - & 1.24&3-5&8.32\\
M 3-26 & -12.63 & -12.05 & 8.6&-13.28&0.40 & 1.18 & 2.47 & 1.01& 7 &7.44\\
M 3-32 & -12.26 & -11.69 & 6.0&-12.77&0.39 & 1.93 & 1.28 & 0.93& 5 &6.90\\
M 3-33 & -12.37 & -11.94 & 5.0&-12.94&0.29 & 1.25 & - & 1.07& 6 &8.52\\
M 3-37 & -13.84 & -12.16 & 10.0&-13.46&1.14 & 0.85 & 0.94 & 0.99& 5 &9.97\\
M 3-52 & -14.40 & -12.39 & 11.6&-13.76&1.36 & 0.35 & 0.9 & -& 5 &10.70\\
M 3-54 & -13.24 & -11.86 & -&-&0.93 & 1.40 & - & 1.09& 7 &8.87\\
M 4-4 & -14.09 & -11.63 & 6.6&-12.75&1.66 & 0.96 & 0.9 & -& 7 &8.33\\
M 4-6 & -14.03 & -11.00 & -&-&2.05 & 5.81 & 1.29 & 1.07& 5 &6.53\\
M 4-7 & -14.18 & -11.68 & -&-&1.69 & 1.58 & 1.29 & 1.41& 5 &5.08\\
M 4-8 & -13.47 & -11.22 & -&-&1.52 & 7.33 & 1.05 & -& 0.5&9.32\\
Pe 1-12 & -13.30 & -12.98 & 9.6&-14.26&0.22 & 9.58 & 1.35 & 1.35& - &8.82\\
Pe 1-13 & -12.96 & -12.32 & 7.6&-13.50&0.43 & - & - & 1.58& 9 &11.42\\
Th 3-13 & -14.10 & -10.95 & -&-&2.14 & 2.52 & 2.22 & -& 2 &10.31\\
Th 3-15 & -13.75 & -12.04 & -&-&1.16 & 0.60 & - & 1.2& 2 &10.16\\
Th 4-2 & -13.57 & -12.81 & 19.0&-14.39&0.52 & 0.39 & 0.85 & 1.02& 7 &11.29\\
Th 4-7 & -13.88 & -12.40 & 6.0&-13.48&1.00 & 2.11 & 1.16 & 1.43& $>$4&5.56\\
Vy 2-1 & -12.03 & -12.03 & 7.0&-13.11&0.00 & 0.84 & 0.94 & 0.73& 3 &5.58\\
\noalign{\smallskip}
\hline\noalign{\smallskip}
H 1-13 & -13.62 & -10.34 & 9.6&-11.62&2.22 & 3.73 & 1.08 & 0.83& 5 &1.43\\
H 1-42 & -11.82 & -10.96 & 5.8&-12.03&0.58 & 2.41 & 1.28 & 0.96& 4 &4.85\\
H 1-51 & -13.83 & -12.87 & 13.2&-14.30&0.65 & 0.24 & 0.79 & 1.57& 5 &-\\
K 6-3 & -14.65 & -11.31 & -&-&2.26 & 1.76 & 0.75 & 1.31& 1 &-\\
K 6-5 & -15.13 & -11.53 & -&-&2.44 & 1.15 & 1.00 & -& - &-\\
M 1-31 & -12.63 & -10.86 &-&-& 1.20 & 3.96 & 1.11 & 0.75& 3 &4.50\\
M 3-9 & -13.43 & -11.74 & 17.0&-13.27&1.15 & 1.96 & - & 1.03& 6 &3.57\\
Th 3-34 & -15.18 & -11.04 & -&-&2.81 & 5.29 & 2.53 & 2.03& 9 &-\\
YC 2-32 & -12.33 & -11.98 &15.0&-13.45& 0.24 & 3.89 & - & 0.94& 5 &-\\
\noalign{\smallskip}
\hline
\end{tabular}
\end{flushleft}
\end{table*}

\subsection{Ionic abundances}

In order to calculate abundances for the ions N$^{+}$, O$^{0}$,
O$^{+}$, O$^{++}$, S$^{+}$, S$^{++}$, Ar$^{++}$, Ne$^{++}$,
we have used the fits by Alexander \& Balick (1997) that show
a good agreement with ionic abundances obtained from
statistical balance equations. For the helium ions (He$^{+}$, He$^{++}$)
we used the recombination coefficients from P\'equignot et al.
(1991), with the coefficients for He$^{+}$ collisional excitation correction
derived by Kingdon \& Ferland (1995).
Only for Cl$^{++}$, not present in the Alexander \& Balick (1997) tables,
we used the statistical balance equation for the [ClIII]$\lambda\lambda$5518+5538 lines.

In particular, the O$^{+}$ abundance can be derived from
two pairs of lines: $\lambda$3727+29 and $\lambda$7319+30.
There is however a systematic difference
between the resulting values, with some tendency to higher abundances
for the red pair. These differences are shown in figure 2. In this figure,
the dashed line is a linear fit for our sample.
To explain this difference, observed
also by other authors, we raised four alternatives:
interstellar extinction, data reduction, electron temperature
and density variations, and O$^{++}$ recombination.

- Interstellar Extinction: one of the factors that could cause
this difference is a poor interstellar
extinction correction. To check this, we corrected all spectra
with two interstellar extinction laws (Fitzpatrick 1999 and CCM 1989)
using two extreme values for R$_{V}$: 2.1 e 5.9.
The results do not imply a significant variation of the O$^{++}$
abundance in any case. The only possibility to invoke extinction
to explain such a difference would be a new, totally distinct interstellar
extinction law, which seems unrealistic, since
most of the extinction comes from the disk

- Data reduction: as the two line pairs are in opposite regions of
the spectrum, poor flux calibration could affect the final data.
However, it is unlikely that this effect could affect all data.
Furthermore, this effect
appears also in other, independent works (Stasi\'nska et al. 1998).

- Temperature and density variations: this can be an important
factor in abundance calculations (Viegas \& Clegg 1994).
Regions with high density in the inner part of a planetary nebula
contribute to the intensity of $\lambda$7319+30, but not to $\lambda$3727
(Mathis et al. 1998). With respect to the density, as the
intensity of the [SII] line relative to H$\beta$ depends on the density
(Alexader \& Balick 1997), this value will represent preferentially
lower density regions, and this effect would be more evident
for bulge objects due to their smaller angular sizes.
For disk objects, the slit would cover only the central part
of a planetary nebula and the effect is not as important as for
bulge nebulae.

- Recombination of O$^{++}$: For nebulae with low density and electron
temperature, dielectronic recombination of O$^{++}$ may play an important
role (Rubin 1986,  Aller \& Keyes 1987), leading to differences in the
derived ionic abundances.

Figure 5 displays a comparison between the O$^+$ ionic abundance derived
from the blue (3726+29) and the red (7320+30)lines, both for our sample
and that from Stasinska et al. (1998). In the figure, the continuous
line is a y=x plot and the dashed line is a linear fit for all the
points. As can be seen, there is a small discrepacy between both
determinations, with a tendency for higher abundances when the red
lines are used. The same effect appears for both samples.
As the final explanation is still an open issue, we adopted the blue
pair to derive the O$^{+}$ abundance like most of the other works in the
literature:

$$ log\rm{(O^{+}_{3727}/H^{+})} = (-0.1\pm0.3)+(1.06\pm0.08) log\rm{(O^{+}_{7325}/H^{+})} $$

\bigskip

\begin{figure}[h]
\epsfxsize=230pt \epsfbox{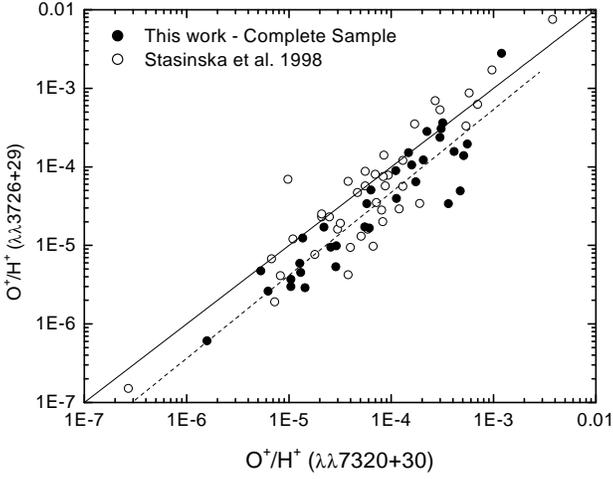}
\caption[]{A comparison of the O$^{+}$ ionic abundance from
the blue (3727+29) and red (7319+30) lines, respectively.}
\end{figure}

The derivation of sulfur abundance requires both S$^+$ and S$^{++}$ ions. For
those objects where S$^{++}$ was not available we adopted the same technique
used by Kingsburgh \& Barlow (1994), deriving the S$^{++}$ abundance from 
the relation between ratios S$^{++}$/S$^{+}$
and O$^{++}$/O$^{+}$. This relation was derived from our data, as
can be seen in figure 6. In the figure, filled and open circles represent the points 
included in the fit, and crosses are abundances poorly determined, which were
not included. The dotted line is a y=x relation
and the continuous line is the best fit represented by the relation:

$$ log\rm{(S^{++}/S^{+})} = (0.21\pm0.08)+(0.73\pm0.06) log\rm{(O^{++}/O^{+})} $$

\bigskip

\begin{figure}[h]
\epsfxsize=230pt \epsfbox{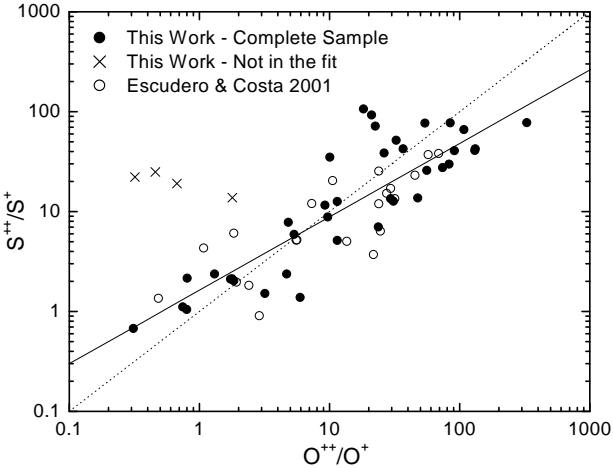}
\caption[]{Relation between S$^{++}$/S$^{+}$ and O$^{++}$/O$^{+}$}  
\end{figure}

\subsection{Elemental abundances}

The final abundances are listed in table 4. The objects separated
by an horizontal line in the final of table 4 and also in table 3 are those from our
Sample 2 (see section 2.1). In the discussion on the abundance distribution
(see section 5) we do not include these objects.
 
Since it is not possible to get the chemical abundances of all ions of
a given element, we use the ICF method (Ionization
Corrections Factors) to derive the elemental abundances,
as in our previous works. For details, see Escudero \& Costa
(2001), Costa et al. (1996) and references therein.

The adopted ICF for oxygen and nitrogen are those suggested by Peimbert
\& Torres-Peimbert (1977); for sulfur and neon we used the ICF provided
by Kingsburgh \& Barlow (1994), except for those objects for which S$^{++}$ 
is not available; in this case we adopted the value
derived by the relation given in the last section. For argon we used the ICF provided by
Freitas Pacheco et al. (1993). These ICFs are listed below.

$$ \frac {\rm{He}} {\rm{H}} = \frac {\rm{He^{+}}} {\rm{H^{+}}} + \frac {\rm{He^{++}}} {\rm{H^{+}}} $$
$$ \frac {\rm{O}} {\rm{H}} = (\frac {\rm{O^{++}}} {\rm{H^{+}}} + \frac {\rm{O^{+}}} {\rm{H^{+}}})
\frac {\rm{He}} {\rm{He^{+}}} $$
$$ \frac {\rm{N}} {\rm{H}} = \frac {\rm{N^{+}}} {\rm{H^{+}}} \frac {\rm{O}} {\rm{O^{+}}} $$
$$ \frac {\rm{S}} {\rm{H}} = (\frac {\rm{S^{+}}} {\rm{H^{+}}} + \frac {\rm{S^{++}}} {\rm{H^{+}}})
(1 - (1- \frac {\rm{O^{+}}} {\rm{O}})^{3})^{-1/3} $$
$$ \frac {\rm{Ar}} {\rm{H}} = 1.34 \frac {\rm{Ar^{++}}} {\rm{H^{+}}}
\frac {\rm{O}} {\rm{O^{++}}} $$
$$ \frac {\rm{Ne}} {\rm{H}} = \frac {\rm{Ne^{++}}} {\rm{H^{+}}} \frac {\rm{O}} {\rm{O^{++}}} $$

\begin{table}
\caption[]{Chemical Abundances}
\begin{flushleft}
\begin{tabular}{lllllll}
\hline\noalign{\smallskip}
Name & He & $\epsilon$(N) & $\epsilon$(O) & $\epsilon$(S) &
$\epsilon$(Ar) & $\epsilon$(Ne)\\
\noalign{\smallskip}
\hline\noalign{\smallskip}
Al 2-O & 0.160 & 8.43 & 8.66 & 6.57 & 6.78 & 7.95\\
Bl 3-13 & 0.105 & 7.68 & 8.91 & 7.14 & 6.54 & 8.06\\
H 1-11 & 0.106 & 8.02 & 8.67 & 6.87 & 6.28 & 7.95\\
H 1-14 & 0.103 & 7.62 & 8.27 & 6.39 & 5.66 & 7.38\\
H 1-30 & 0.147 & 8.78 & 8.76 & 6.66 & 6.86 & 8.13\\
H 1-34 & 0.013 & 8.14 & 8.45 & 7.04 & 6.34 & -\\
H 1-41 & 0.123 & 7.95 & 8.64 & 6.81 & 6.22 & 7.85\\
H 1-43 & 0.011 & 8.44 & 8.84 & 6.80 & - & -\\
H 1-45 & 0.099 & 7.39 & 7.52 & 5.69 & 4.97 & 6.84\\
H 1-46 & 0.096 & 7.93 & 8.35 & 6.95 & 6.16 & 7.40\\
H 1-63 & 0.059 & 7.48 & 8.06 & 6.86 & 6.09 & 6.97\\
H 1-67 & 0.125 & 8.50 & 8.75 & 7.09 & 6.64 & 8.00\\
H 2-11 & 0.151 & 8.16 & 8.29 & 6.56 & 6.64 & -\\
H 2-25 & 0.081 & 7.56 & 8.26 & 6.74 & 6.70 & -\\
H 2-48 & 0.048 & 7.12 & 7.93 & 6.41 & 6.03 & -\\
He 2-260 & 0.009 & 7.14 & 8.18 & 6.29 & 6.97 & -\\
He 2-262 & 0.118 & 7.82 & 8.48 & 6.66 & 6.24 & 7.55\\
KFL 2 & 0.113 & - & 8.24 & - & 5.99 & 7.33\\
KFL 4 & 0.099 & - & 8.34 & - & 5.71 & 7.28\\
KFL 19 & 0.080 & 7.38 & 8.65 & 6.93 & 6.06 & 7.64\\
M 1-48 & 0.143 & 8.65 & 8.84 & 7.09 & 6.88& 8.33\\
M 2-7 & 0.137 & 7.91 & 8.74 & 6.61 & 6.80 & 7.91\\
M 2-8 & 0.223 & 8.50 & 8.54 & 6.04 & 6.40 & 7.81\\
M 2-12 & 0.010 & 7.96 & 8.56 & 6.60 & 7.86 & -\\
M 2-13 & 0.117 & 8.22 & 8.72 & 6.96 & 6.48 & 8.07\\
M 2-15 & 0.128 & 8.30 & 8.61 & 6.95 & 6.38 & 8.00\\
M 2-20 & 0.081 & 8.07 & 8.78 & 6.83 & 6.43 & -\\
M 2-26 & 0.161 & 8.77 & 8.49 & 6.88 & 6.65 & 7.53\\
M 2-38 & 0.152 & 8.57 & 8.58 & 7.28 & 6.73 & 7.78\\
M 2-39 & 0.107 & 6.94 & 7.48 & 5.45 & 5.64 & 6.76\\
M 3-19 & 0.101 & 7.27 & 8.00 & - & 6.29 & 7.22\\
M 3-26 & 0.137 & 8.43 & 8.40 & 6.99 & 6.36 & 7.92\\
M 3-32 & 0.132 & 8.27 & 8.56 & 6.95 & 6.35 & 7.87\\
M 3-33 & 0.072 & 7.72 & 8.62 & 6.57 & 6.00 & 7.84\\
M 3-37 & 0.092 & 8.34 & 8.86 & 6.72 & 6.74 & 8.34\\
M 3-52 & 0.129 & 8.58 & 8.79 & 7.15 & 6.91 & -\\
M 3-54 & 0.122 & 7.33 & 8.67 & 6.52 & 6.24 & 7.97\\
M 4-4 & 0.154 & 8.74 & 8.92 & 7.25 & 6.78 & -\\
M 4-6 & 0.111 & 8.47 & 8.71 & 6.98 & 6.30 & 8.01\\
M 4-7 & 0.102 & 7.94 & 8.33 & 6.43 & 5.79 & 7.59\\
M 4-8 & 0.015 & 7.24 & 8.03 & 6.24 & 6.66 & -\\
Pe 1-12 & 0.114 & 7.90 & 8.63 & 5.41 & 6.36 & 7.82\\
Pe 1-13 & 0.099 & - & 9.03 & - & 6.66 & 8.29\\
Th 3-13 & 0.089 & 7.33 & 7.61 & 6.10 & 5.98 & -\\
Th 3-15 & 0.107 & 7.62 & 8.06 & 6.15 & 5.89 & -\\
Th 4-2 & 0.180 & 8.72 & 8.50 & 6.95 & 6.75 & 7.79\\
Th 4-7 & 0.072 & 7.70 & 8.35 & 6.47 & 5.90 & 7.49\\
Vy 2-1 & 0.145 & 9.04 & 8.94 & 6.23 & 6.94 & 8.10\\
\noalign{\smallskip}
\hline\noalign{\smallskip}
H 1-13 & 0.123 & 8.84 & 8.99 & 7.47 & 6.67 & 8.20\\
H 1-42 & 0.106 & 8.27 & 8.69 & 6.97 & 6.15 & 7.94\\
H 1-51 & 0.098 & 8.27 & 9.51 & 6.94 & 7.72 & 9.38\\
K 6-3 & 0.128 & 7.98 & 8.33 & 6.40 & 6.49 & -\\
K 6-5 & 0.159 & 9.12 & 8.57 & 7.32 & 6.88 & -\\
M 1-31 & 0.161 & 9.02 & 8.98 & 7.69 & 6.89 & 8.28\\
M 3-9 & 0.059 & 7.69 & 8.68 & 5.19 & 6.34 & 8.05\\
Th 3-34 & 0.193 & 9.15 & 8.14 & 7.08 & 6.43 & -\\
Y-C 2-32 & 0.088 & 7.59 & 8.59 & 6.61 & 6.19 & 7.81\\
\noalign{\smallskip}
\hline
\end{tabular}
\end{flushleft}
\end{table}

\section{Errors in the physical parameters and chemical abundances}

In Escudero \& Costa (2001) typical errors were derived through
a Monte Carlo simulation, based on the line intensities and their
errors, which requires the same exposure time for all measurements.
This is not the case for the present work, so we examined the
effect of these uncertainties on the abundances.

From Escudero \& Costa (2001), we have estimated average
errors to the abundances of 0.023, 0.18, 0.17, 0.24 and 0.19, 
respectively for He, N, O, S and Ar.
However, it should be remembered that errors are highly affected
by interstellar extinction, in the sense that
objects that have higher extinction should have greater errors
in the abundances, on the average, depending also on the exposition time
and the size of the telescope. Therefore, abundances of bulge objects, that have
high extinction, typically have higher errors than disk objects, so that
we can estimate the uncertainties in the abundances of bulge nebulae to be
roughly 0.2 dex, apart from He/H.

\subsection{Physical parameters}

One of the main error sources in the chemical abundances are the uncertainties
in electron temperatures and densities, as discussed previously. Low S/N
in weaker diagnostic lines, as well as internal variations in the physical
parameters certainly affect the final abundances. As discussed in section
3.2, electron densities derived from [SII] can be underestimated since the
emissivity of this line depends on density.

\subsection {Helium}

The main factors that affect the helium abundance are the uncertainties in densities
and electron temperatures. Depending on these parameters, the correction due
to collisional excitation can reach 50\%. Therefore, errors in the physical
parameters would considerably affect the chemical abundances.
The existence of neutral helium would also have an influence on
 the derived abundances.
This was tested on the basis of the criterion proposed by Torres-Peimbert \& Peimbert
(1977), according to which objects with log O$^+$/O $>$ -0.4 seem to have a substantial
contribution of neutral helium. For this sample, 11 objects fit this criterion,
indicating that their derived helium abundance should be taken cautiously.

\subsection {Abundances from collisionally excited lines}

In addition to observational errors and those resulting from the inhomogeneity
of the nebulae, the use of ionization correction factors (ICF) to estimate elemental
abundances for the elements with collisionally excited lines will include
additional error sources.

For nitrogen, O$^{+}$ is used as ICF, and small variations in its abundance
will affect the nitrogen abundance. The use of O$^{+}_{3727}$ or O$^{+}_{7325}$
ionic abundances
will imply differences of 0.0 to 0.7 dex in the nitrogen abundance.

The same effect occurs for sulfur, in whose ICF O$^{+}$ is also present.
Two observational effects affect the sulfur abundance: the dependence
of S$^{+}$ emissivity with density, and the weak intensity of the line
[SIII]$\lambda6312$, which will result in a poor determination of the
S$^{++}$ abundance. Therefore, the present sulfur abundances should be taken
cautiously.

For oxygen, argon and neon, whose lines are normally intense in planetary
nebulae, the main error sources are the uncertainties in the physical
parameters.

\section {Data from the literature}

To compare our data, we collected objects from 5 recent works in the
literature: Stasi\'nska et al. (1998),
Cuisinier et al. (2000), Cuisinier et al.
(1996), Samland et al. (1992) and K\"oppen et al. (1991). We use
only those objects with distances already determined
(Schneider \& Buckley 1996, van de Steene \& Zijlstra 1995
and Zhang 1995), and with heliocentric distances greater than 5 kpc.

In figure 7 we show the distribution of chemical abundances of He, N and O
from the literature and our data. 

The results from the first paper
of this series are also shown (solid irregular lines). 
In order to homogenize the results, we rederived the elemental abundances
for those data, using the same ionic abundances but adopting the improvements
used here: sulfur ICF from Kingsburgh \& Barlow (1994); O$^+$ abundance
derived preferentially from the blue doublet, and S$^{++}$ derived acording
the procedure described in section 3.2, when necessary.

These results have
been obtained using the same instruments and procedures as in the present paper,
so that both samples are expected to present a large degree of homogeneity.
This fact can be observed in fig. 7, as the abundance distributions
of both samples are similar. Small differences observed at the low end of
the He and N abundances can be attributed to the incompleteness of the samples.
It can be seen that our abundances show a good
similarity to all other data. The nitrogen abundances, that reflect the
progenitor mass, show a little more scattering from our objects than the data
from literature. This was expected to occur because our objects are more
widely distributed
in angular position than the literature objects (see Escudero \& Costa 2001).

\begin{figure}[h]
\epsfxsize=230pt \epsfbox{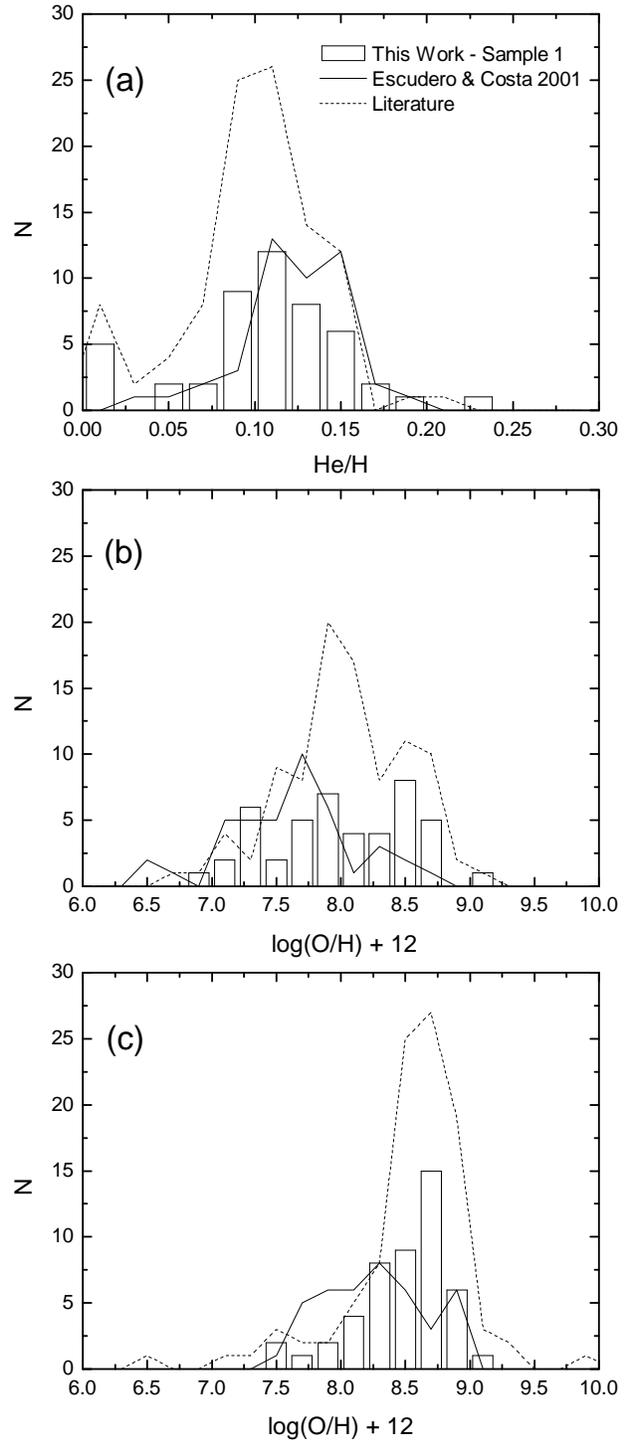}
\caption[]{Distribution of helium, nitrogen and oxygen abundances for 
this work (Sample 1), from Escudero \& Costa (2001) and from selected sources in the
recent literature (see text).}
\end{figure}

\section {Discussion and conclusions}

Our results are in agreement with those from other works
(see for example Escudero \& Costa 2001, Cuisinier et al. 2000),
indicating that bulge PNe have mean abundances similar to
those from disk objects, however displaying higher dispersion.

Figure 8 displays the relation between excitation classes and helium abundances
(panel a) and oxygen abundances (panel b). The data distribution in panel (a)
shows no correlation between the helium abundances and the excitation
class; the only exception is that some objects with very low abundances
are preferably associated with the lowest excitation classes, which possibly
indicates the presence of neutral helium for these objects.
On the other hand, the absence
of correlation in panel (b) also indicates that our abundance analysis 
is unbiased with respect to the excitation of the nebulae.

\begin{figure}[h]
\epsfxsize=230pt \epsfbox{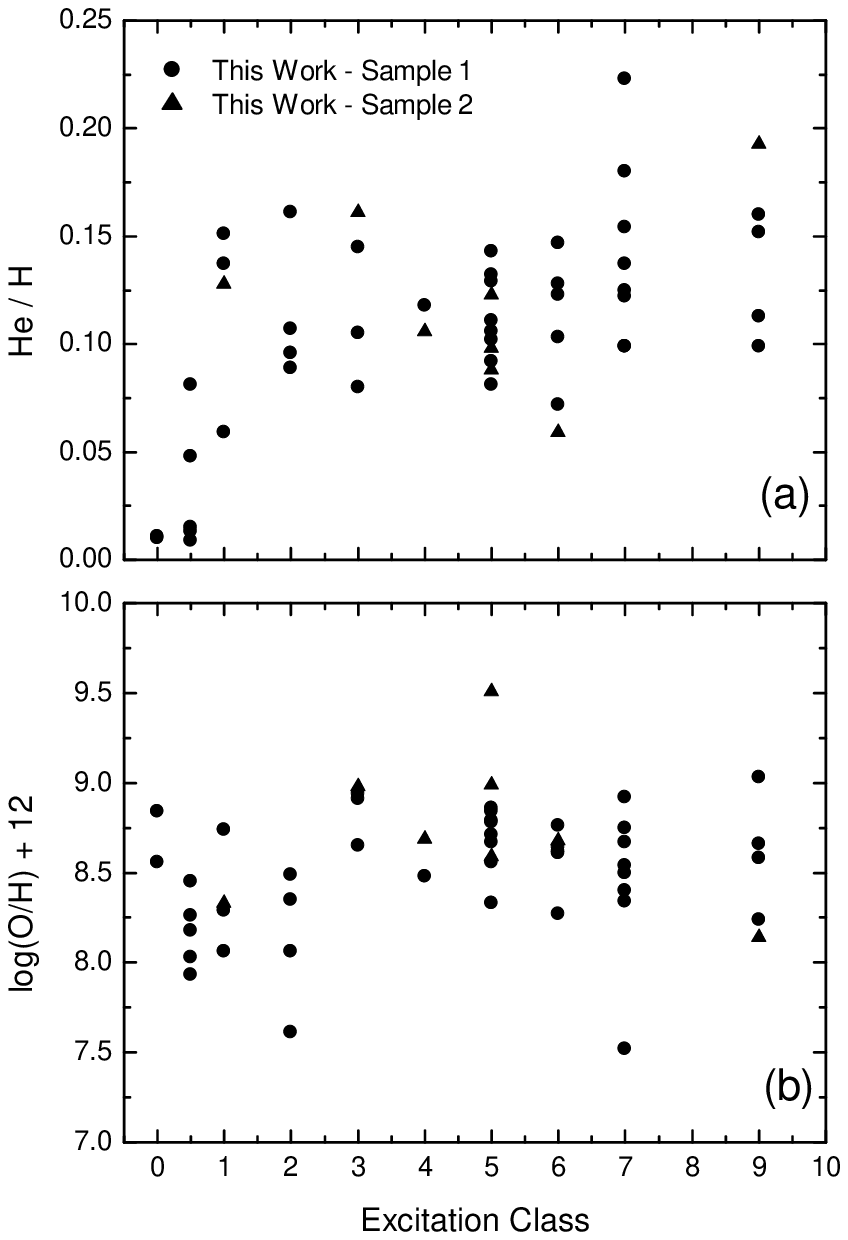}
\caption[]{Excitation classes versus helium (a) and oxygen (b) abundances.}
\end{figure}

Figure 9 displays several abundance correlations for our sample which are
distance independent, combined with
data from Escudero \& Costa (2001) and the literature.

Oxygen and nitrogen are two key elements for the diagnostics
of PNe. Due to strong lines, oxygen abundances are normally very well
determined, giving the abundance of the interstellar medium at
the progenitor formation epoch. It was already noted that
more massive, younger objects like type-I PNe progenitors have
oxygen abundance lower than the Sun or type-IIa PNe progenitors
(see Costa et al. 1996). The same effect appears in stellar evolution
yields (van den Hoek \& Groenewegen 1997, Marigo 2001), however
for the majority of the lower mass stars, oxygen remains unchanged
and can be used to trace the chemical evolution of the interstellar
medium.

\begin{figure*}
\centering
\includegraphics[width=17cm]{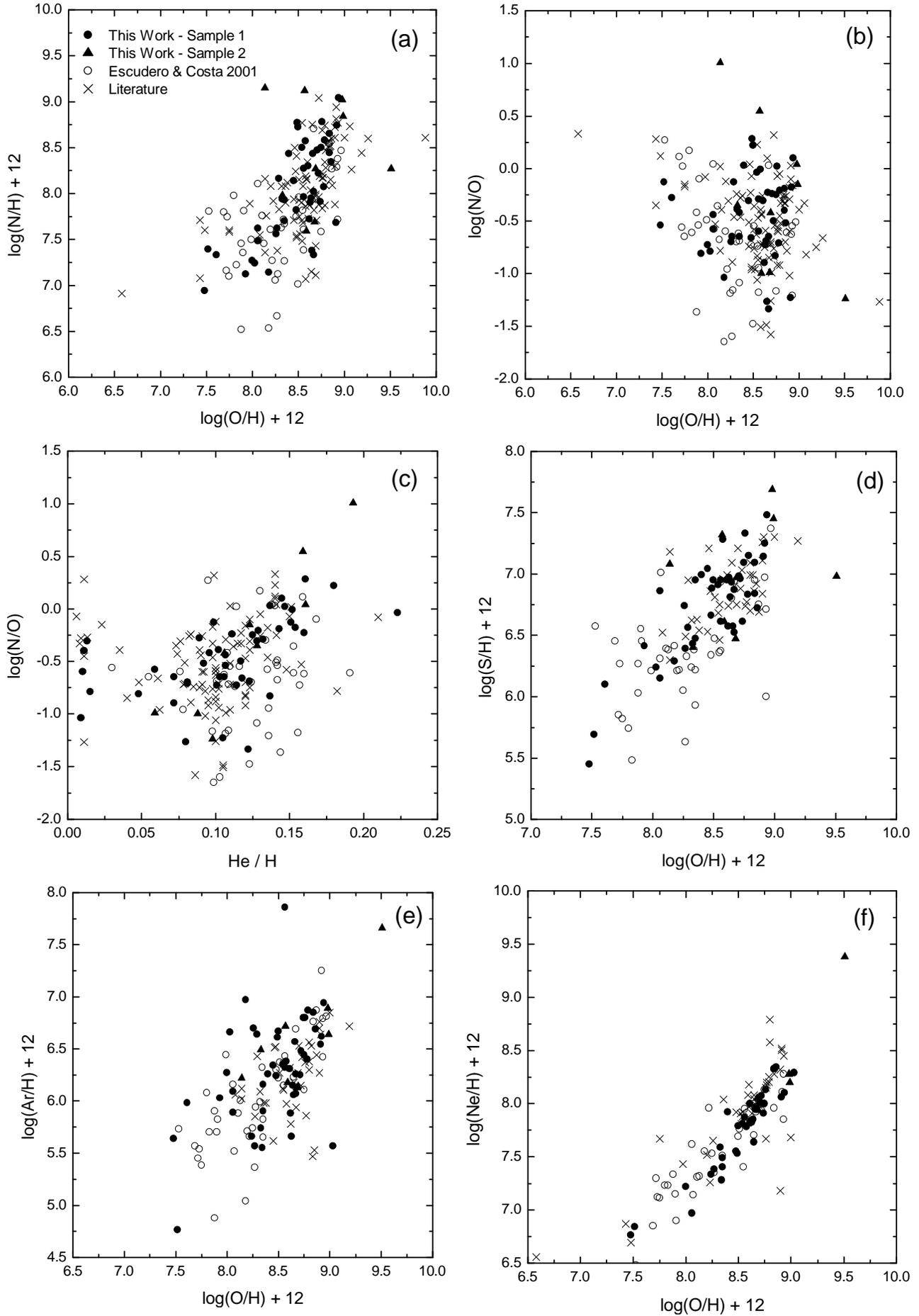}
\caption[]{Abundance correlations from our results and the literature(see section 6).}
\end{figure*}

Nitrogen, on the other hand, has two possible origins: it reflects the abundance
of the interstellar medium at the progenitor formation epoch, combined with
the material produced by nucleosynthesis during the evolution
of the PN progenitor star. Typical yields as a function of the stellar mass are
given by van den Hoek \& Groenewegen (1997) and Marigo (2001).

Examining the O vs. N diagram displayed in fig. 9a,
a positive correlation between O/H and N/H can be observed. 
As for the correlation between log(N/O) and O/H (figure 9b),
this is less evident.
We can see that objects with high N/O can have higher or lower oxygen abundances.
Those with higher  oxygen abundances are similar to the disk type-I PNe, while
those with lower abundances are similar to halo objects.
On the other hand, objects with low N/O are usually oxygen-rich, which is
consistent with the correlation shown in fig. 9a. Some of these objects
may be type-III PNe in the inner disk
(see also Maciel 1999) and with the abundances of metal rich bulge
giants (Rich 1988).
This dispersion in the abundances suggests that the chemical evolution
of the bulge cannot be explained in
 a single formation scenario, so that a composite scenario is required instead.

Another important diagram for the diagnostics of PNe abundances is
He x log(N/O), which reflects the chemical enrichment during
the stellar evolution of the progenitor. From figure 9c
it can be seen that there is a reasonably well defined correlation
between He/H and log(N/O), in the sense that helium-rich nebulae are
are usually also nitrogen rich, which probably reflects
an increasing progenitor mass. A small group of objects apparently
have very low He abundances, while their N abundances are normal. This
can be explained either by the fact that their progenitor star had
smaller masses, which characterizes them as type IIa nebulae, or their
He abundances may have been underestimated owing to the presence of
neutral helium.

Figures 9d, 9e and 9f display the correlation between the alpha-elements;
as they are produced mainly in type II supernovae their correlation reflects
the chemical evolution of the interstellar medium from which the progenitor
stars are formed. From these figures one can see that lower oxygen abundances
imply also lower sulfur, argon and neon abundances.

\begin{acknowledgements}
We thank E.L. Fitzpatrick for providing his extintion curves. We thank also
A. Acker, referee of this work, for her comments and suggestions.
This work was partly supported by the Brazilian agencies \emph{FAPESP} and
\emph{CNPq}. A.V.E. acknowledges \emph{FAPESP} for his graduate fellowship (Process 00/12609-0).
Observations at ESO/Chile were possible through the \emph{FAPESP} grant 98/10138-8.
\end{acknowledgements}

\end{document}